\documentclass{elsart}
\usepackage{epsfig}
\usepackage{amssymb}
\newcommand{\msun}{\mbox{$M_\odot$}}

\def\be{\begin{eqnarray}}
\def\ee{\end{eqnarray}}
\def\lsim{\mathrel{\rlap{\lower3pt\hbox{\hskip1pt$\sim$}}
     \raise1pt\hbox{$<$}}} %less than or approx. symbol
\def\gsim{\mathrel{\rlap{\lower3pt\hbox{\hskip1pt$\sim$}}
     \raise1pt\hbox{$>$}}} %greater than or approx. symbol

\def\bi{\bibitem}

\begin{document}

\runauthor{Brown, Lee, Rho}

\begin{frontmatter}
\title{Relation of Strangeness Nuggets to Strangeness Condensation and the
Maximum Mass of Neutron Stars}

\author[suny]{Gerald E. Brown,}
\author[pnu,apctp]{Chang-Hwan Lee,}
\author[pnu]{Hong-Jo Park,}
\author[saclay]{and Mannque Rho}

\address[suny]{Department of Physics and Astronomy,\\
               State University of New York, Stony Brook, NY 11794, USA \\
(\small E-mail: Ellen.Popenoe@sunysb.edu)}

\address[pnu]{Department of Physics, Pusan National University,
              Pusan 609-735, Korea\\
          (E-mail: clee@pusan.ac.kr, hongjopark@pusan.ac.kr) }
\address[apctp]{Asia Pacific Center for Theoretical Physics,
POSTECH, Pohang 790-784, Korea}

\address[saclay]{Service de Physique Th\'eorique,
 CEA Saclay, 91191 Gif-sur-Yvette
c\'edex, France
(E-mail: rho@spht.saclay.cea.fr)}

%\thanks[geb]{Ellen.Popenoe@sunysb.edu}
%\thanks[chl]{clee@pusan.ac.kr}
%\thanks[rho]{rho@spht.saclay.cea.fr}

\renewcommand{\thefootnote}{\fnsymbol{footnote}}
\setcounter{footnote}{0}

\begin{abstract}
The recent experimental indications of density dependence in the
pion decay constant $f_\pi^\star$ and $\omega$ meson mass
$m_\omega^\star$ and the discovery of $S^0 (3115)$ and other
strange ``nuggets" are providing a strong support for kaon
condensation in dense hadronic matter, thereby re-kindling the
interest in the issue of the critical stable mass of neutron
stars. The density-dependent quantities provide increases in the
vector mean fields mediated by $\rho$ and $\omega$-meson exchange
which increase by a factor $\sim 1.56$ the Weinberg-Tomozawa term
in kaon-nucleon interactions which accounts for $\sim$ half of the
binding energy of the $K^-$ meson in dense matter. Furthermore
lattice gauge calculations have pinned down the value of $KN$
sigma term, $\Sigma_{KN}$, the explicit chiral symmetry breaking
in the strangeness sector. The partial rotation out of this
explicit breaking provides the other $\sim$ half of the $K^-$
binding energy. The net result is to confirm the work of Thorsson
et al. that strangeness condensation takes place at $u=n/n_0
\simeq 3$, where $n_0$ is nuclear matter density, in neutron
stars. We suggest in this article that a support for this scenario
is provided by the recent experiments of Suzuki et al. who found
tightly bound strangeness nuggets. The strangeness nuggets
discovered in the experiments involve approximately the same ratio
of nucleons to $K^-$ meson as in the center region of neutron
stars. But whereas the latter can be described by mean fields, in
which the medium effects are substantially more attractive than in
the finite system, the binding in neutron star matter should be
substantially (say, $\sim$ 20\%) greater than that in the
strangeness nugget. This would strengthen the argument by Brown
and Bethe that the accompanying softening in the equation of state
should limit the maximum neutron star mass to $M_{\rm NS}^{\rm
max} \simeq 1.5\msun$. This low $M_{\rm NS}^{\rm max}\sim
1.5\msun$ has major consequences in astrophysics, especially for
the merging rate of compact stellar objects.
\end{abstract}
%\begin{keyword}
%\PACS{97.60.Lf; 97.80.Jp}
%\end{keyword}

\end{frontmatter}

\renewcommand{\thefootnote}{\arabic{footnote}}
\setcounter{footnote}{0}
%%%%%%%%%%%%%%%%%%%%%%%%%%%%%%%%%%%%%%%%%%%%%%%%%%%%%%%%%%%%%%%%%%%%%%%%%%%%%%%%
\section{Introduction\label{intro}}

Based on the detailed numerical calculations of Thorsson et al.
\cite{TPL1994}, Brown and Bethe \cite{BB1994} claimed that the
result of kaon condensation setting in at a density $n\sim 3 n_0$,
where $n_0$ is nuclear matter density, gives a sufficient
softening of the equation of state in neutron star matter to limit
the maximum possible neutron star mass to $\sim 1.5\msun$. In the
analysis of Supernova 1987A, which Bethe and Brown considered to
have evolved into a black hole, these authors \cite{BB1995} were
able to establish a maximum mass for the compact object of
$1.56\msun$, based on the observed amount of nickel produced in
the supernova explosion. The $M_{\rm NS}^{\rm max}\sim 1.5\msun$
was consistent with 1987A evolving into a low-mass black hole.
Furthermore, the ZAMS (Zero Age Main Sequence) $18\msun$
progenitor of 1987A was calculated to have an Fe core mass of
$\sim 1.5\msun$. Brown et al.\cite{Brown2001,BWW} argued that this
Fe core mass was about the same as the compact object mass on the
basis of calculations by Woosley showing that fallback in the
supernova explosion compensated for the increased binding energy
of the compact object.

We claim that the above astrophysical phenomenon involving neutron
star masses is intricately connected with the softness in the
equation of state coming when the electrons in the neutron star
can shed their high degeneracy energy by changing into
$K^-$-mesons, in a zero momentum Bose condensate. This is a
particle and nuclear physics problem which in our view must
incorporate the vacuum structure of dense matter involving chiral
symmetry and hence in-medium properties of hadronic masses. Our
focus here will therefore be on particle/nuclear physics aspect.
We shall make a brief summary of the astrophysical consideration
in a section below but leave the detailed discussion to a longer,
technical paper \cite{BBL2006}.

The standard chiral perturbation approaches predominantly employed
in the field are anchored on chiral Lagrangians whose parameters
are determined in the matter-free vacuum (that we shall refer to
as ``zero-vacuum") and computations are done by perturbation
around the zero-vacuum in terms of a set of presumed small
expansion parameters. Our approach will differ from them in a
crucial way. Ours will be a mean-field approach with a Lagrangian
defined in a sliding vacuum~\cite{BR2004}. The basic idea is
similar to Landau Fermi-liquid theory for many-body systems as
applied to nuclear matter~\cite{BR2002}. Here one performs
``double decimation"~\cite{BR2004}. The crucial point is that {\it
the mean field approximation with an effective chiral Lagrangian
with scaling parameters is equivalent to Fermi-liquid fixed point
theory.} It is not known how to obtain the effective Lagrangian
with ``sliding vacua" from an effective Lagrangian defined in the
zero-vacuum just as it is unknown how to derive from a fundamental
chiral Lagrangian the four-Fermi interaction Lagranian effective
near the Fermi surface that via Wilsonian RGE gives rise to
Fermi-liquid fixed point theory~\cite{shankar}. In the same vein,
our starting Lagrangian that is defined with ``sliding vacua" is
not derived in any systematic way from a zero-vacuum chiral
Lagrangian. It will be the same Lagrangian that gives rise to the
Fermi-liquid fixed point theory, here extended to three flavors.
We will employ mean field approximation with this Lagrangian. Kaon
condensation will emerge as instability against strangeness
condensation like superconductivity resulting from an unstable
Fermi surface triggered by the Cooper pairing. Since in this way
of approaching kaon condensation one is approaching from below the
QCD phase transition, namely, chiral restoration, we will call
this ``bottom-up" approach.

Since kaon condensation must take place before, and in the
vicinity of, the chiral restoration point, a mean field approach
should make a better sense if one fluctuates around the
``vector-manifestation (VM)" fixed point discovered by Harada and
Yamawaki~\cite{HY:PR} of hidden local symmetry Lagrangian
Wilsonian matched to QCD. Within the framework of hidden local
symmetry (HLS) approach to hadron physics, there are two points at
which the theory is reliably known. One is the zero-vacuum around
which fluctuations can be computed with confidence if vector
mesons are treated on the same footing as the pions. A suitable
perturbation expansion treating the vector meson mass as a small
parameter in the sense of the $1/N_c$ expansion is found to give
results in good agreement with experiments in matter-free
space~\cite {HY:PR}. The other is the vector manifestation (VM)
point at which in the chiral limit, the light-quark vector mesons
become massless and the parameters of the Lagrangian controlling
low-energy physics get fixed to a definite value and to which HLS
flows as temperature, density or the number of flavors is dialled
to the critical value for chiral restoration. If one is not too
far from the VM point, it can be substantially advantageous to
start from the VM fixed point at which the Lagrangian is known. We
will call this ``top-down" approach. An extreme case where this
approach is successful even for a process taking place in
matter-free space is the chiral doubler splitting of the D meson
discussed in \cite{HRS}~\footnote{It is perhaps significant to
note that in this process, even though the starting point is the
VM, the tree contribution dominates, loop corrections making up
only 1/3 of the total.}. Other examples are discussed in
\cite{mr-dsb04} where it has been suggested that certain processes
in baryonic environment and/or particularly sensitive to the
presence of vector degrees of freedom are much more effectively
treated starting from the VM than from the zero-vacuum. This
``top-down" approach is not yet fully formulated for the process
in question. So we can only give a drastically simplified
treatment below

In mean-field in the bottom-up approach on which we will be mainly
focused, there are two main driving forces towards kaon
condensation. One is the movement towards the restoration of the
explicitly broken chiral symmetry $\Sigma_{KN}$ in the strangeness
sector~\cite{kaonrot}. This is characterized in the calculations
by the chiral symmetry parameter $a_3 m_s$ related to the $KN$
sigma term. We shall update the calculations of strangeness
condensation here of Thorsson et al. \cite{TPL1994} who used three
different values of $a_3 m_s$. This parameter has now been
calculated on the lattice \cite{Dong1996}, with central value $a_3
m_s = -231$ MeV, and quoted accuracy of $3-4\%$ which is only
slightly greater in magnitude than the Thorsson et al. central
value of $-222$ MeV which we shall use. Furthermore, Thorsson et
al. used nuclear compression moduli of $K_0=180$ and 240 MeV. We
favor the value of 210 MeV for which $M_{\rm NS}^{\rm
max}=1.5\msun$. Kaon condensation sets in at $n=3.08 n_0$, where
$n$ is the density and $n_0$ is nuclear matter density and the
maximum neutron star mass is $1.5\msun$.

The second main driving force towards kaon condensation comes from
the Weinberg-Tomozawa term treated in the mean field. This comes
from vector mean fields between the nucleons and the $K^-$.
Following \cite{BRwalecka}, in mean field, the force mediated by
the $\omega$-meson exchange between a $K^-$ and a nucleon can be
related to that between two nucleons as
 \be
V_{K^-}(\omega) = -\frac 13 V_N \label{eq1}
 \ee where $V_N$ is the
vector mean field from nucleons in nuclear matter. The 1/3 is easy
to understand, because there is only one nonstrange antiquark in
the $K^-$, whereas there are three nonstrange quarks in the
nucleon. The $\rho$ meson exchange gives
 \be V_{K^-} (\rho) =\pm \frac
13 V_{K^-} (\omega)
 \ee repulsive for neutrons, attractive for protons. In the
Weinberg-Tomozawa term, the vector-meson propagator for symmetric
nuclear matter is
approximated by zero momentum,
 \be V_{K^-}(\omega) = - \frac{3}{8 F_\pi^2} n,\label{3}
 \ee
where $n$ is the nucleon density and $x_p$ ($x_n$) is the proton
(neutron) fraction, the connection with the mean fields being by
way of a KSRF-type relation
 \be m_V^2 = 2 F_\pi^2 g_V^2.\label{4}
 \ee
In eqs.(\ref{3}) and (\ref{4}), $F_\pi$ is a parameter that
appears in the HLS Lagrangian which is related to the pion decay
constant $f_\pi$~\footnote{We make the distinction between $F_\pi$
and $f_\pi$. It is $f_\pi$ which is an order parameter of chiral
symmetry vanishing at the critical point in the chiral limit.
Harada and Yamawaki show that $f_\pi\equiv f_\pi (Q^2=0)=F_\pi +
\Delta$ where $\Delta$ is a quadratically divergent term that
cancels $F_\pi$ at the critical point. Note that it is $F_\pi$
that we are concerned with here.} and $g_V$ is the vector (or
flavor gauge) coupling.

Two recent experimental developments provide crucial information
that allows to determine the above mean field potentials in
medium:
\begin{enumerate}
\item The ``$f_\pi$" connected with the $\rho$-meson via
the KSRF relation has been found in deeply bound pionic atoms to
be substantially below its free space value
\cite{Suzuki2004}~\footnote{We should note here that it is really
$F_\pi$ that is determined in this experiment, not the order
parameter $f_\pi$. However at the mean field level and up to
nuclear matter density, they are the same and hence the
interpretation as ``partial restoration of chiral symmetry" as
measured in nuclei is correct. This aspect is frequently confused
in the literature.},
 \be
\left(F_\pi^\star(n_0)/F_\pi\right)^2 = 0.65 \pm 0.05 \label{eq5}
\ee extracted at tree order. Other determinations are reviewed by
Brown and Rho \cite{BR2004}.
\item As what may be taken as the first $direct$ and unambiguous
verification of the scaling proposed in \cite{BR91}, the decrease
in $\omega$-meson mass with density has been measured as
$m_\omega^\star/m_\omega\simeq 0.84$ at $n=n_0$ \cite{Metag2004}.
This is somewhat larger than the 0.8, consistent with
Eq.~(\ref{eq5}), which we shall use; we believe $f_\pi^\star$ to
be more accurately determined by the extensive
data on deeply bound pionic atoms.
\end{enumerate}

As shown by Harada and Yamawaki~\cite{HY:PR}, $g_V$ changes with
scale, such that both $g_V^\star$ and $m_V^\star$ go to zero
proportionally to the quark condensate $\langle \bar{q}q\rangle$
near the VM fixed point as $g_V^\star/m_V^\star$ constant. In fact
there is some evidence that this behavior sets in from $n\simeq
n_0$ upward~\cite{BR2004,SBMR,Song}. However, there is no evidence
that $g_V^\star$ scales appreciably below $n=n_0$. Brown and Rho
\cite{BR2004} have roughly summarized the situation by letting
$m_V^\star$ scale up to nuclear matter density as
$m_V^\star(n_0)/m_V\simeq 0.8$, with $g_V$ constant. Beyond $n_0$
$g_V^\star/m_V^\star$ is taken to be constant. This is certainly a
rough description, but it reproduces the main features of the
Harada-Yamawaki scaling.

With the scaling in $F_\pi^2$ only up to nuclear matter density
$n_0$, we find the mean fields to be increased by a factor
 \be
\left(\frac{F_\pi}{F_\pi^\star}\right)^2\simeq
\left(\frac{1}{0.8}\right)^2\simeq 1.56.\label{enhance}
 \ee

An astonishing new development has been the experimental discovery
of deeply bound kaonic nuclear states \cite{Suzuki04}. The
interpretation by Akaishi et al. \cite{Akaishi2005} (see, for
related discussions, Mares et al.~\cite{Mares2} who studied the
widths of deeply bound kaonic atoms and arrived at 100 $\sim$ 200
MeV binding energy in these states) is of particular interest in
connection with the Thorsson et al. \cite{TPL1994} calculation and
our present work is bringing it up to date. Although the deeply
bound kaonic-nuclear states are small, essentially the strangeness
(anti)nuggets suggested in many papers, the players in the
nuggets, two neutrons, one proton and a $K^-$-meson, are the same
as those in the Thorsson et al. mean field approach, at the
density beyond kaon condensation where one $K^-$ is present for
every two neutrons. We show that the Thorsson et al. mean field
analysis gives substantial support to the Akaishi et al. analysis.
In return, since medium dependent effects, essentially the
dropping meson masses, have larger effects in the mean field than
in the ``nugget" calculations, experimental confirmation of the
latter would give strong support to the work of Thorsson et al.
and consequently the low $M_{\rm NS}^{\rm max}$ of Brown and Bethe
\cite{BB1994}.

The plan of this note is as follows. In Sec.~\ref{sec2} we shall
schematize the Thorsson et al. (bottom-up) calculation, carefully
separating effects from scalar and vector mean fields in lowest
order, showing how they get mixed up in higher order. We also
discuss the range term, introduced in higher-order in
\cite{LBMR,Lee:PR} and show why the medium dependence in this
effect has confounded investigators who obtain input parameters
from $K^-$-nucleon scattering lengths. We then show in
Sec.~\ref{sec3} how the mean field calculation is related to the
strangeness ``nugget" calculation of Akaishi et al.
\cite{Akaishi2005}. In Sec.~\ref{sec4}, we discuss certain
specific features associated with the VM point when the treatment
is made top-down from the VM. In Sec.~\ref{sec5} we review some of
the many other calculations of strangeness condensation and
discuss why $\Sigma^-$ hyperons do not seem to figure importantly
in dense matter, or low $-$ even if they do $-$ they  would not be
expected to alter our $M_{\rm NS}^{\rm max}$ appreciably.

Although the detailed analysis of the evolution of compact
binaries, double neutron stars and low-mass black-hole,
neutron-star binaries, will be put off to a later paper
\cite{BBL2006}, we point out the obvious consequences of a low
$M_{\rm NS}^{\rm max}$ mass in Sec.~\ref{sec6}; namely that the
giant progenitors of a binary neutron star must be very close
($\sim 4$ \%) in ZAMS mass. As more and more double binary neutron
stars are observed, this can easily be checked. We shall point out
that present observations already defy the statistics of the
standard model of binary neutron star evolution, necessitating an
alternative model.

In Appendix, we construct a crystalized dense system comprised of
tribaryon-$K^-$ bound states. We find that the $K^-$'s get unbound
for densities exceeding $2 n_0$. This indicates that in dense
medium kaon condensation can set in starting from $n\sim 2 n_0$.
Since we started with the fixed proton fraction (1/3) in this
approach, however, this critical density cannot be directly
compared with the critical density for kaon condensation in the
mean field approach.

%%%%%%%%%%%%%%%%%%%%%%%%%%%%%%%%%%%%%%%%%%%%%%%%%%%%%%%%%%%%%%%%%
\section{Strangeness Condensation}
\label{sec2} In this section we sketch the principal arguments for
our bottom-up approach.
\subsection{Scalar channel}
\begin{figure}
\centerline{\epsfig{file=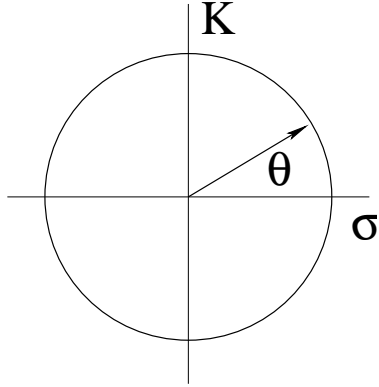,height=2.0in}} \caption{
Projection onto the $\sigma, K$ plane. The angular variable
$\theta$ represents fluctuations toward the kaon mean field. }
\label{fig1}
\end{figure}

We first discuss the contribution towards the restoration of the
explicitly broken chiral symmetry, essentially considering
$V$-spin~\cite{kaonrot} (see Fig.~\ref{fig1}). This way of looking
at kaon condensation does not capture $all$ aspects of the
phenomenon but it gives a clear conceptual picture of what is
involved.

The Hamiltonian for explicit chiral symmetry breaking is
\be
H_{\chi SB} &=& \Sigma_{KN} \langle\bar N N\rangle \cos\theta
+\frac 12 m_K^2 F_\pi^2 \sin^2\theta  \nonumber\\
&\simeq & \Sigma_{KN} \langle\bar N N\rangle \left(1-\frac{\theta^2}{2}\right)
+\frac 12 m_K^2 F_\pi^2 \theta^2 \label{eq7}
 \ee
where the last expression is obtained for small fluctuation
$\theta$ in the direction of the $K$-axis. We then find \be
{m_K^\star}^2 = m_K^2 \left(1-\frac{\Sigma_{KN} (\bar NN)}{f_\pi^2
m_K^2} \right) \label{eq8} \ee showing that the effective kaon
mass $m_K^\star$ is strongly reduced by the explicit chiral
symmetry breaking $\Sigma_{KN}$ in the strangeness sector.

As noted above, the explicit chiral symmetry breaking has been
evaluated in lattice calculations by Dong et al. \cite{Dong1996}
giving $\Sigma_{KN}= 362$  MeV. Although in chiral perturbation
theory the scalar nuclear density $\langle\bar\psi\psi\rangle$ can
be replaced by the vector density
$n=\langle\psi^\dagger\psi\rangle$ since the above expression is
in the highest order calculated, we prefer to keep the scalar
density because at $\sim 3 n_0$ we may have gone over to
constituent quark degrees of freedom. Looking at the behavior of
the order parameter $f_\pi^\star(n)$, we see that chiral symmetry
restoration can only be declared\footnote{Whereas Nambu-Jona
Lasinio theory gives chiral restoration at $n_c\approx  2
n_0$~\cite{Bernard87}, the transition from nucleon to constituent
quark degrees of freedom, which NJL starts with, should take
another $\sim 2 n_0$. With finite temperature the two transitions,
one from nucleon to constituent quarks and then chiral restoration
are discussed in \cite{BGLR}. It is possible to approach also $n_c
\lsim n_0$ from the fixed point (at which $m_\rho^\star
\rightarrow 0$ in the chiral limit) since the role of Brown-Rho
scaling is known ($m_\rho^\star /m_\rho \simeq 1-0.2 n/n_0$) for
$n<n_0$ (see our later eq.~(\ref{eq19})) and then the mass drops
linearly in $\langle\bar q q\rangle^\star$ for higher densities.}
at $n\sim 4 n_0$. Although one cannot take this result at its face
value, we take this to suggest
that %the scalar density $\langle\bar\psi\psi\rangle$
the nucleon effective mass decreases
linearly from $n=0$ to $n_c= 4 n_0$. This gives
$\langle\bar\psi\psi\rangle/\langle\psi^\dagger \psi\rangle$ for
the proton as 0.82,
$\langle\bar\psi\psi\rangle/\langle\psi^\dagger \psi\rangle$ for
neutron as 0.57 when we take 90 \%  neutrons and 10 \%  protons as
Thorsson et al. \cite{TPL1994} did at $n=3 n_0$.
So, the contribution of $\Sigma_{KN}$ will be reduced by about 20\%
if we take the scalar density instead of vector density.
In the arguments given below, instead of taking the scalar density
self-consistently, we use vector density with
smaller $\Sigma_{KN}=310$ MeV, which correspond to $a_3 m_s=-222$ MeV,
for the comparison with the results of Thorsson
et al.\cite{TPL1994}.

Now, one notable development following the Thorsson et al. work
was the calculation of the range term \cite{Lee:PR}
 \be
\Sigma_{KN}^{\rm eff} =\left(1-0.37
\left(\frac{\omega_{K^-}^\star}{m_{K^-}}\right)^2\right)\Sigma_{KN}.
 \ee
For self consistency the final {\it in-medium} $\omega_{K^-}$
should be used here although this is at odds with the standard
chiral counting that we are $not$ adhering to. If we use the
Thorsson et al. $\mu_e=219$ MeV at $n_c=3.08 n_0$ for $a_3 m_s =
-222$ MeV (remember that the $K^-$ mass
$m_{K^-}^\star\equiv\omega_{K^-}^\star$ is equal to $\mu_e$ at the
phase transition.) then
 \be \Sigma_{KN}^{\rm eff} = 0.93
\Sigma_{KN}.
 \ee
Without range term it follows from Eq.~(\ref{eq8}) that
 \be m_K^\star=333\; {\rm MeV},
 \ee
whereas with $\Sigma_{KN}^{\rm eff} =0.93 \Sigma_{KN}$ \be
m_K^\star=347\; {\rm MeV}. \label{eq12} \ee Thus, with
introduction of the range term there would be a 14 MeV correction
upwards in the $\mu_e$ necessary for kaon condensation in Thorsson
et al.\cite{TPL1994}.

The use of the full $\Sigma_{KN}$ is highly preferable to the
procedure used by most research workers who obtained
$\Sigma_{KN}^{\rm eff}$ from fitting the $K^-$-nucleon scattering
lengths at threshold. They would have obtained \be
\Sigma_{KN}^{\rm eff} = 0.63 \Sigma_{KN} \;\;\; ({\rm
threshold\;\; scattering}) \ee since $\omega_{K^-}=m_{K^-}$ is
assumed in this procedure.~\footnote{Indeed, for the pion the
range term is $\sim -1.1 \Sigma_{\pi N}$ and its introduction
changes a quite appreciable attractive interaction into a slightly
repulsive one. The fact that the range term in pion scattering is
so well known makes it inexcusable to omit it in the $K^-$-nucleon
scattering.}

\subsection{Vector channel}
We now turn to the Weinberg-Tomozawa term. As noted in
Eq.~(\ref{eq1}), for $\omega$-exchange the mean field is $
V_{K^-} (\omega) = - (1/3) V_N$, and for the
$\rho$-meson
$V_{K^-}(\rho) = \pm (1/3) V_{K^-}(\omega)$.
%%\footnote{This expression is not consistent with Eq.~(\ref{3}).}

We calculate first with these mean fields. With the 90 \% neutrons
and 10 \% protons at $n_c=3.1 n_0$, the vector mean field
contribution to $\omega_{K^-}$ would be \be V_{K^-}(\omega)=
-\frac 13 \frac{g_\omega^2}{m_\omega^2} \frac 12
\left(\frac{x_n}{2}+x_p\right) n \simeq -126 \; {\rm MeV }, \ee
where $g_\omega = 3 g_\rho$ with $g_\rho \sim 5$, and $x_{n,p}$
are the neutron and proton fractions. Putting this together with
the $m_K^\star=347$ MeV of Eq.~(\ref{eq12}) gives us \be
\omega_{K^-}^\star (n=3 n_0) = 221\; {\rm MeV}, \label{eq14} \ee
essentially the same as Thorsson et al. \cite{TPL1994}. However,
from the medium dependence in hand, we increase the magnitude of
the 126 MeV by 1.56 given by (\ref{enhance}), giving an additional
$\sim 70$ MeV drop in the $K^-$ mass. As noted, 14 MeV of this is
used up in the range term, so we are left with 56 MeV more binding
than Thorsson et al. \cite{TPL1994}.

That the enhancement factor (\ref{enhance}) is called for is
clearly indicated in the structure of nuclear matter. Were we to
take the free pion decay constant $f_\pi \simeq 93$ MeV in the
$K^-$ vector potential (\ref{3}), it would give,  at nuclear
matter density,
 \be
 V_{K^-} =\frac 13 V_N \simeq 58 \; {\rm MeV}
  \ee
which would mean that the $\omega$-mean field for the nucleon
would be $V\simeq 174$ MeV. This is much weaker than the Walecka
mean fields which are employed at nuclear matter density $n=n_0$.
They are more like $V({\rm Walecka}) \gsim 270$ MeV; i.e., $\gsim
50\%$ higher. Thus, the phenomenology favors the mean field
growing with density.

In fact, the vector mean field is proportional to
$g_V^2/m_\omega^2$ and a recent experimental study shows that
$m_\omega^\star$ decreases with density \cite{Metag2004}. Much
more study has been made of the density dependence of
$f_\pi^\star$ in deeply bound pionic atoms which arrive at
 \be
\left(\frac{F_\pi^\star (n_0)}{F_\pi}\right)^2=0.65 \pm 0.05
 \ee
extracted at tree level \cite{Suzuki2004}.

A similar
 \be
\frac{F_\pi^\star (n_0)}{F_\pi}\simeq 0.8
 \ee
can be obtained \cite{Lutz} from the Gell-Mann-Oakes-Renner
relation~\footnote{Note that here and in the next two equations,
there is no difference between $f_\pi$ and $F_\pi$ since we are
going up to $n\simeq n_0$.}
 \be
f_\pi^2 m_\pi^2 = -\hat m \langle \bar q q\rangle
 \ee
under the assumption that $m_\pi$ and $\hat m$ do not scale with
density, giving $\left(f_\pi^\star/f_\pi\right)^2=\langle\bar q
q\rangle^\star/ \langle\bar q q\rangle$, and then using the model
independent
 \be
\frac{\langle\bar q q\rangle^\star}{\langle\bar q
q\rangle} =1-\frac{\Sigma_{\pi N} n}{f_\pi^2 m_\pi^2}.
\label{eq19}
 \ee
Thus by now we have ample evidence of scaling in $m_\rho^\star$
and $m_\omega^\star$.~\footnote{The parametric masses of the
$\omega$ and $\rho$ mesons should scale in the same way in medium
in HLS theory if one assumes $U(N_f)$ symmetry but the pole masses
can differ a bit due to medium-dependent loop corrections. Here we
are focusing on the former since we are working in the mean field
approximation.}  By taking $f_\pi^\star/f_\pi\simeq 0.8$ in our
formulae we assume them to drop from their bare value by 20 \% at
nuclear matter density $n_0$. This gives our factor 1.56 increase
in the vector mean fields.

%As noted before, Harada and Yamawaki \cite{HY:PR} show that
%$g_V^2/m_V^2$ must scale in the same way going into the fixed
%point for $g_V$ and $m_\rho$, both going to zero (in the chiral
%limit) at the fixed point. We will mention below empirical
%evidence of the vector meson decoupling at higher momentum. Brown
%and Rho \cite{BR2005} handle this in a rough way by assuming only
%$m_V$ to scale up to $n=n_0$, and then $g_V^\star/m_V^\star$ to
%remain constant up to the fixed point, essentially what was found
%empirically in \cite{SBMR,Song}.

%%%%%%%%%%%%%%%%%%%%%%%%%%%%%%%%%%%%%%%%%%%%%%%%%%%%%%%%%%%%%%%%%
\section{(Anti) Strange Nugget}
\label{sec3}

Recent experiments \cite{Suzuki04,Suzuki05} and theoretical
interpretation \cite{Akaishi2005} involve the same players, e.g.
two neutrons, one proton and a $K^-$ in the strange tribaryon
$S^0(3115)$ as in the neutron star matter. For example, in Table~4
of Thorsson et al. \cite{TPL1994} which refers to $a_3 m_s =- 222$
MeV for $n\sim 3.8 n_0$ the proportion of $K^-$ is $x_{K^-} \sim
0.5$, the $K^-$ having replaced the electrons, so that the ratio
of protons to neutrons is $\sim 0.5$ for overall charge
neutrality. For this value of $x$ the system is well into kaon
condensation (with threshold $u=n/n_0=3.08$; note the similarity
to mean density of $S^0 (3115)$ of $u=3.1$). Thus, if one cuts out
a piece of the strangeness condensed star (which is on its way
into a black hole) that contains two neutrons, it will have one
proton and one $K^-$.

First of all, in our description, the star should undergo
strangeness condensation at a lower value of $u$ than the nugget.
In the description of the star, mean fields are used; e.g., the
Weinberg-Tomozawa term is
 \be V_{\rm ave} = -\frac 13
\frac{g_V^2}{m_V^2} n,
 \ee
for the above composition of two neutrons for each proton. Note
that since $V_{\rm ave}$ is a mean field, it is evaluated for zero
momentum transfer. This maximizes the medium effect
in terms of the dropping vector meson mass $m_V \rightarrow
m_V^\star$.

In fact, Brown et al. \cite{BLS} have argued that in the GSI
experiments with nucleon and kaon momenta $|P_N|\sim 444$ MeV and
$|P_K|\sim 322$ MeV, form factors of $f_V(p)\sim 0.82$ must be
employed in order to take into account the finite size of the
nucleon. There is a partial decoupling of the vector interaction.
Thus, the interactions in the nugget should be somewhat weaker
than in the infinite system, both because of the decoupling of the
vector interaction and because medium effects are maximum for mean
fields.

The same interactions enter into the neutron star and strangeness
nugget calculations, but the neutron star is well and truly
strangeness-condensed, well beyond threshold for the composition
at which the nugget is formed. This seems reasonable in terms of
the stronger mean field interactions.

Akaishi et al. \cite{Akaishi2005} formulate their strangeness nugget
problem in terms of potentials. Whereas the Weinberg-Tomozawa term,
which is a vector potential, gives the main attraction between the
two neutrons and the proton, we prefer to formulate the other main
attraction as coming from the partial restoration of the explicitly
broken chiral symmetry in the strangeness sector, essentially a scalar
mean field. This would require something like a local density approximation
in $\langle\bar\psi\psi\rangle$ in the nugget problem, which would be
of doubtful accuracy.

We agree with Akaishi et al. \cite{Akaishi2005} that the $\sim 50
$ MeV difference between the binding energy (143 MeV) they
obtained and the observed one (194 MeV) comes from the medium
effects (``Similar to the case of observed pionic bound states"
\cite{Akaishi2005} as we discussed in $f_\pi$ being replaced by
$f_\pi^\star$) and it should be noted that in our description --
which is different in spirit from theirs, we get about the same
$\sim 50$ MeV as Akaishi et al.

In other words, we believe that there is a mapping of the neutron-star
problem onto the strangeness nugget problem, or vice versa, and
that the discovery of the nugget strongly supports the former.

%%%%%%%%%%%%%%%%%%%%%%%%%%%%%%%%%%%%%%%%%%%%%%%%%%%%%%%%%%%%%%%%%
\section{${\mathbf a=1}$ At and Away From the Fixed Point}
\label{sec4}

The behavior of the various quantities we work with should be
better determined at the fixed point of Harada and Yamawaki
\cite{HY:PR} where $m_\rho^\star$ and $g_V^\star$ both go to zero,
linearly with $\langle \bar q q\rangle^\star$ which goes to zero
(as does $f_\pi^\star$) and the quantity $a$, which we now define
$\rightarrow 1$. Mean field approximations are expected to get
better the closer we are to the VM fixed point~\cite{HY:PR}. This
suggests the top-down approach to kaon condensation as well as to
the strangeness nugget problem. Here we present in the simplified
form what we can say about kaon condensation.

The low energy theorem of the hidden gauge theory valid to all
orders of chiral perturbation theory~\cite{HY:PR} is
 \be
{m_\rho^\star}^2 =a^\star F_\pi^\star {g_V^\star}^2,
 \ee
defining $a$ in terms of the other variables we use. Earlier, when
we talked about integrating out the $\rho$-mass, we used the
KSRF-type relation
 \be m_\rho^2 = 2 F_\pi^2 g_V^2
 \ee
which follows from vector dominance in free space, $a$ being then
equal to 2 at the scale of $m_\rho$. However, in turns out
\cite{HY:PRL} that the vector dominance at $a=2$ is on an unstable
trajectory of RG flow of the HLS theory with no connection to the
trajectory that leads to the Harada and Yamawaki vector
manifestation and that the fact that in nature the vector
dominance model seems to work in matter free space is merely an
accident. In fact, apart from the pionic form factor at zero
temperature and zero density, $a$ near 1 provides a highly
satisfactory phenomenology. In particular, $a=1$ provides a good
description for the coupling of the photon to the
nucleon.~\cite{BR2004}. Other evidences for $a\simeq 1$ in nature
including chiral doubling in $D$ mesons are discussed in
\cite{mr-dsb04}.~\footnote{It is shown in \cite{HY:PR} that at the
matching scale $\Lambda\sim 4\pi f_\pi\sim 1$ GeV, HLS Lagrangian
is given by $a\simeq 1.3$. Since this Lagrangian is expected to
hold in the large $N_c$ limit, one can think of $a\simeq 1.3$ as
representing the large $N_c$ limit. It turns out however that as
far as phenomenology is concerned, $a\simeq 1$ gives as good a fit
as $a\simeq 1.3$.} Thus, we find it more appropriate to use
 \be
\frac{{g_V^\star}^2}{{m_\rho^\star}^2} =\frac{1}{a^\star
{F_\pi^\star}^2} \label{eq21}
 \ee
with $a^\star$ not far from 1, as compared with the
 \be \frac{{g_V}^2}{{m_\rho}^2} =\frac{1}{2
{F_\pi}^2} \label{eq22}
 \ee
which pertained to matter-free space in which the vector meson was
integrated out to get the Weinberg-Tomozawa relation. As can be
seen, the vector mean field interaction is thus found to be
increased by a factor of $2/a^\star$, or a factor of 2 as the
fixed point is reached.

Our next objective is to estimate (in the chiral limit) at which
density $n_c$ the fixed point is reached. We do this by finding
out at which density $m_\rho^\star$ goes to zero.

Although initially $m_\rho^\star$ decreases as $\sqrt{\langle \bar
q q\rangle^\star}$ following the scaling of $F_\pi^\star$ (see
argument above following Eq.~(\ref{eq19})) the Harada and Yamawaki
work shows that once it starts dropping it scales as $\langle \bar
q q\rangle^\star$. (See the empirical verification of this in Koch
and Brown \cite{KochBrown93} who showed that the entropy matched
that in LGS if the meson masses were allowed to scale as
$\langle\bar q q\rangle^\star$, which was referred to as ``Nambu
scaling.") As noted earlier, $g_V$ does not seem to scale up to
nuclear matter density $n_0$, but then Nambu scaling sets in.
Nambu scaling is $\sqrt 2$ times faster than the initial scaling
of $m_\rho^\star$ from $n=0$ to $\sim n_0$, which decreases
$m_\rho^\star$ by 20\%. Thus, we believe in the interval $n_0$ to
$2 n_0$, $m_\rho^\star$ will decrease $\sqrt 2$ times 20\%, or
$\sim 28\%$, and the same from $2 n_0$ to $3 n_0$, and from $3
n_0$ to nearly $4 n_0$ where $m_\rho^\star=0$ in the chiral limit.
Thus, the fixed point at $n_c$ is at $n\lsim 4 n_0$.

{}From our earlier argument that $g_V^\star$ scales as
$m_\rho^\star$ for $n > n_0$, but up to $n_0$, $g_V$ remains
constant, whereas $m_\rho^\star$ scales, we find that
Eq.~(\ref{eq21}) can be expressed as \be
\frac{{g_V^\star}^2}{{m_\rho^\star}^2} =\frac{1}{a^\star} \left(
\frac{1}{0.8 F_\pi}\right)^2 \ee and we know that $a^\star=1$ at
$n_c$, where it, together with $m_\rho^\star$ and $g_V^\star$, has
a fixed point. Compared with the matter-free expression
Eq.~(\ref{eq22}), which is the KSRF relation, we see that
 \be
\frac{[{g_V^\star}^2/{m_\rho^\star}^2]_{\rm fixed \; point}}
     {[{g_V}^2/{m_\rho}^2]_{\rm zero \; density}}
     \simeq\frac{2}{0.8^2} \simeq 3.13.
 \ee
Thus, the mean field felt by the $K^-$ is increased by the factor
3.13 when the scaling of both $F_\pi^\star$ and $a$ are included,
at the fixed point at $n_c$, the final doubling coming from the
scaling in $a^\star$. At the VM fixed point, the condensate is
zero (in the chiral limit), so the scalar contribution from the
rotation of $\Sigma_{KN}$ is gone. Thus seen from the VM point,
the essential doubling of the attractive vector interaction
replaces the attraction given by the rotating out of the
$\Sigma_{KN}$ term. In fact within the range of $a$ relevant to
the problem~\cite{mr-dsb04}, say, $1\lsim a\lsim 1.3$, the
coefficient of the density $n$ needed to bring the $m_{K^-}^\star$
down, is relatively constant up to $n_c$. And furthermore the
scenario is close to the bottom-up scenario discussed above.

We emphasize that our above argument connects strangeness
condensation with chiral restoration, in that with the large
increase in vector mean fields in going to the fixed point,
strangeness condensation certainly takes place at an $n$ below
$n_c$.

Following Eq.~(\ref{eq14}) it was noted that because of medium
effects, we had an additional 56 MeV binding of the $K^-$ as
compared with Thorsson et al. Rather than a
$\mu_{e,c}=m_{K^-}^\star (n_c)\simeq 219$ MeV of these authors, we
would find 221 $-$ 56 = 165 MeV.

Now in our calculation about the fixed point, with the same 90\%
neutrons and 10\% protons at $n_c$, we would have the factor of
3.13 times 126 MeV giving 394 MeV binding, just 100 MeV in
magnitude less than the kaon mass in matter-free space. However,
if we took $a=1.3$ as suggested by large $N_c$ considerations, we
would have 302 MeV binding, with $\mu_{e,c} = m_{K^-}^\star = 192$
MeV. Thus, within the possible range of $a$ that could be arrived
at by the RG flow from the fixed point (we believe $\simeq 1.3$ to
be maximal), we find the $\mu_{e,c}$ of Thorsson et al., corrected
by medium effect, to be midway in the range of $\mu_{e,c}$'s that
could be reached starting from the fixed point. A similar
attraction was obtained for $n_P/n\simeq 0.25$ by Tsushima et
al.~\cite{tsushima} in a different approach that takes into
account the scaling behavior.

We see from Thorsson et al., for $a=1-1.3$, the necessary electron
chemical potential is easily reachable by $n=3 n_0$ so strangeness
condensation seems assured by $3 n_0$. In fact, the mass of the
Hulse-Taylor pulsar is $1.44\msun$ and since it must be
stabilized, it will give a lower bound on $n_c$.

The Kaplan-Nelson term involving $\Sigma_{KN}$ that is important
in the bottom-up approach to kaon condensation in an infinite
matter would be difficult to include reliably in the strangeness
nugget calculation, so it would be far more advantageous to carry
it out top-down from the VM point. Since they do not deal with
mean fields, but rather with relatively high vector meson momenta
in
 \be
\frac{{g_V^\star}^2}{{m_V^\star}^2+q^2}
 \ee
they will not benefit as much as the kaon condensation calculation
from the change in $a$ from 2 in free space to $\sim 1$ in medium.
However, the factor 1.56 for $(F_\pi/F_\pi^\star)^2$ remains.
Thus, we might expect the binding there to be $\sim 1.56 \times
126$ MeV =147 MeV. In fact 126 MeV was the original vector mean
field for 90\% neutrons and 10\% protons whereas Akaishi et al.
have two neutrons for each proton, so the number would be 160 MeV
at $n=3 n_0$. Hence the binding would be $1.56\times 160$ MeV =
250 MeV.

%%%%%%%%%%%%%%%%%%%%%%%%%%%%%%%%%%%%%%%%%%%%%%%%%%%%%%%%%%%%%%%%%
\section{The Lack of $\Sigma^-$ Role}
\label{sec5}

The mass of the $\Sigma^- pn$ system lies at $\sim 3075$ MeV, well
below that of the $S^0(3115)$, and one might think that this
system plays a role in strangelets. Similarly, in neutron stars,
the $\Sigma^-$ will replace both a nucleon and an electron. This
consideration led many investigators to consider $\Sigma^-$
condensation, most recently by Kolomeitsev and Voskresensky
\cite{KV2003}.

We do not think that the $\Sigma^-$ obstructs our principal
argument. Batty et al.\cite{BFG} noted that ``for $K^-$ atoms, the
fitted potential becomes repulsive inside the nucleons, implying
that $\Sigma$ hyperons generally do not bind in nuclei." One can
see from Fig.~4 in Brown et al. \cite{BLRapp1998} that in the
density dependencies favored by Batty et al. \cite{BFG}, the
$\Sigma^-$ optical potentials are highly repulsive. Furthermore
there are now direct experimental nuclear data (to be
distinguished from atomic data) that show that the repulsion
increases with the neutron excess \cite{Saha,Noumi}. A DWIA
analysis \cite{Harada} of these ($\pi^-$, $K^+$) data confirms the
repulsion which was first predicted from $\Sigma^-$ atoms by Batty
et al. \cite{BFG94} and extended by Mares et al. \cite{Mares} in a
relativistic mean field approach.

In Kolomeitsev and Voskresensky \cite{KV2003}, although the S-wave
$K^-$ condensate occurs only at $\rho_c\gsim 4 \rho_0$ in neutron
star matter, it is preceded by a long, essentially mixed phase of
hyperon condensation which begins at $u\sim 2$. During a mixed
phase the pressure is constant; even with Gibbs construction the
gradient in pressure is low, so that the force, given by gradient
in pressure, is low. Thus, the neutron star should compress
substantially in this phase and it seems unlikely that the maximum
mass will be greater than our $1.5\msun$.

%%%%%%%%%%%%%%%%%%%%%%%%%%%%%%%%%%%%%%%%%%%%%%%%%%%%%%%%%%%%%%%%%
\section{Astrophysical Consequences}
\label{sec6} In this Section we briefly touch on astrophysical
implications relegating details to a future
publication~\cite{BBL2006}. The important astrophysical
consequence of the low-mass $M_{\rm NS}^{\rm max}$ is that the
standard scenario for binary neutron star evolution \cite{vdH1993}
does not result in a double neutron star, but in a black-hole,
neutron-star binary. In this scenario, after the first born
neutron star is formed, it goes into common envelope evolution
with the companion giant as the latter expands in red giant stage.
During this common envelope evolution it accretes a substantial
amount of matter from the hydrogen envelope of the giant
companion, as it removes this envelope. Bethe and Brown
\cite{BB1998} estimated this amount to be $\sim 1\msun$ for a
$1.4\msun$ neutron star, whereas with more accurate calculation
\cite{BKB} found $\sim 3/4 \msun$ for this mass neutron star. The
accretion is $\sim 0.5\msun$ for a $1.1-1.2\msun$ neutron star.
Obviously these will be sent into black holes and the result will
be a black-hole, neutron star binary.

As shown in \cite{Brown1995}, there is a special way in which the
neutron star evolution in common envelope could be avoided. If the
two giant progenitors are $\lsim 4\%$ different in mass, they will
expand in red giant and burn helium at the same time. Their
hydrogen envelopes are then removed in the double helium star
common envelope evolution. There is no time for the hydrogen to
cross the helium molecular weight barrier \cite{BraunLanger} so
the helium star remains very close in mass. The result is that the
two resulting neutron stars have very nearly equal masses.

\begin{table}
\caption{Compilation of the compact objects in binaries by Lattimer
and Prakash (2004). References are given in their paper.
We have added, following the comma, the recent measurement of
Van der Meer et al.\cite{MKKH}.}
\label{tab1}
\begin{center}
\begin{tabular}{llll}
\hline
Object      & Mass ($\msun$) \phantom{xxxxxx} & Object    & Mass ($\msun$) \\
\hline
\multicolumn{4}{l}{\it X-ray Binaries} \\
4U1700$-$37 & 2.44$^{+0.27}_{-0.27}$ &
Vela X-1    & 1.86$^{+0.16}_{-0.16}$\\
Cyg X-1     & 1.78$^{+0.23}_{-0.23}$ &
4U1538$-$52 & 0.96$^{+0.19}_{-0.16}$ \\
SMC X-1     & 1.17$^{+0.16}_{-0.16}$, 1.05$\pm 0.09$ &
XTE J2123$-$058 & 1.53$^{+0.30}_{-0.42}$ \\
LMC X-4     & 1.47$^{+0.22}_{-0.19}$, 1.31$\pm 0.14$ &
Her X-1     & 1.47$^{+0.12}_{-0.18}$ \\
Cen X-3     & 1.09$^{+0.30}_{-0.26}$, 1.24$\pm 0.24$ &
2A 1822$-$371   & $> 0.73$ \\ %0.97$^{+0.24}_{-0.24}$ \\
\multicolumn{4}{l}{\it Neutron Star - Neutron Star Binaries} \\
1518$+$49           & 1.56$^{+0.13}_{-0.44}$ &
1518$+$49 companion & 1.05$^{+0.45}_{-0.11}$\\
1534$+$12           & 1.3332$^{+0.0010}_{-0.0010}$ &
1534$+$12 companion & 1.3452$^{+0.0010}_{-0.0010}$ \\
1913$+$16           & 1.4408$^{+0.0003}_{-0.0003}$ &
1913$+$16 companion & 1.3873$^{+0.0003}_{-0.0003}$ \\
2127$+$11C           & 1.349$^{+0.040}_{-0.040}$ &
2127$+$11C companion & 1.363$^{+0.040}_{-0.040}$ \\
J0737$-$3039A        & 1.337$^{+0.005}_{-0.005}$ &
J0737$-$3039B        & 1.250$^{+0.005}_{-0.005}$ \\
J1756$-$2251         & 1.40$^{+0.02}_{-0.03}$ &
J1756$-$2251 companion & 1.18$^{+0.03}_{-0.02}$ \\
\multicolumn{4}{l}{\it Neutron Star - White Dwarf Binaries} \\
B2303$+$46  & 1.38$^{+0.06}_{-0.10}$ &
J1012$+$5307 & 1.68$^{+0.22}_{-0.22}$ \\
J1713$+$0747 & 1.54$^{+0.007}_{-0.008}$ &
B1802$-$07   & 1.26$^{+0.08}_{-0.17}$ \\
B1855$+$09   & 1.57$^{+0.12}_{-0.11}$ &
J0621$+$1002 & 1.70$^{+0.32}_{-0.29}$ \\
J0751$+$1807 & 2.20$^{+0.20}_{-0.20}$ &
J0437$-$4715 & 1.58$^{+0.18}_{-0.18}$ \\
J1141$-$6545 & 1.30$^{+0.02}_{-0.02}$ &
J1045$-$4509 & $<$ 1.48 \\
J1804$-$2718 & $<$ 1.70 &
J2019$+$2425 & $<$ 1.51 \\
\multicolumn{4}{l}{\it Neutron Star - Main Sequence Binaries} \\
J0045$-$7319 & 1.58$^{+0.34}_{-0.34}$ \\
\hline
\end{tabular}
\end{center}
\end{table}

In the Table~\ref{tab1}
we show the presently measured double neutron star binary masses.
In the Hulse-Taylor pulsar 1913$+$16 the pulsar mass is $1.44\msun$,
the most massive neutron star in the double neutron star binaries.
It has been evolved from an $\sim 20\msun$ giant \cite{Burrows}.
Since the number of giants go as
\be
N(M) \propto \frac{1}{M^{2.35}}
\ee
according to the Salpeter mass function, if there were no correlation
in pulsar and companion masses, its companion would be most likely to
come from an $M\sim 10\msun$ giant, with neutron star mass
$\sim 1.2\msun$.

As can be seen from the accurately measured 1534$+$12, 1913$+$16,
and 2127$+$11C there is a close correlation in the masses of the pulsar
and companion.

In the case of the lower-mass binaries, the double pulsar J0737$-$3039A,B
and J1756$-$221 equality of masses has been disturbed by the helium star
evolution. In the chain of evolutions, the first born neutron star
has a helium star companion, and low-mass helium stars expand in their
own red giant stage. In so doing they transfer $\sim 0.1 -0.2\msun$
to the neutron star, but since these are of low mass, this is not enough
to send them into a black hole.

Since substantial mass is transferred in the neutron-star, hydrogen-envelope
common envelope evolution for the more massive stars, the above correlation
in neutron star masses within a binary does not give an accurate limit
on $m_{\rm NS}^{\rm max}$. However, Brown's \cite{Brown1995}
special double helium star
scenario is an order of magnitude less likely than the standard scenario,
giving $\sim 10$ times fewer double neutron star binaries. This translates
into 10 times more black-hole, neutron-star binaries than double neutron-star
binaries, which, with the higher masses of the former because of accretion,
should give a factor of $\sim 20$ increase in the gravitational waves
from merging to be detected by LIGO \cite{BB1998}.

Since Table~\ref{tab1} contains three masses, those of 4U 1700$-$37,
Vela X-1 and J0751$+$1807 which exceed our $1.5\msun$ maximum neutron
star mass, we should comment briefly.

{\it 4U 1700$-$37} :
Although this compact object has the same accretion history as the
other high-mass X-ray binaries, it doesn't pulse like the others.
Brown, Weingartner and Wijers \cite{BWW} evolve the compact object
as a low-mass black hole.

{\it Vela X-1} : J. van Paradijs et al.~ \cite{paradijs2} pointed
out that in this binary with floppy B-star companion, the apparent
velocity can in some cases increase by up to 30\% (from the
surface elements of the companion swinging around faster than the
center of mass) ``thereby increasing the apparent mass of the
compact object by approximately the same amount". In any case,
Barziv et al.\cite{barziv}  from which the Vela X-1 neutron star
mass in our table comes, say ``The best value of the mass of Vela
X-1 is $1.86\msun$. Unfortunately, no firm constraints on the
equation of state are possible, since systematic deviations in the
radial-velocity curve do not allow us to exclude a mass around
$1.4\msun$ as found for the other neutron stars."

{\it J0751$+$1807} : We are unable to give a critical discussion
of this binary because thus far only preliminary results have been
published. The sum of neutron-star and white-dwarf mass are
determined by the period change from gravitational radiation.
There is an indication of a Shapiro shift and a white dwarf mass
of $0.188\pm 0.012\msun$ is arrived at. Details of this are not
given (of course it could be obtained from the measurement of the
Shapiro shift, but none is given.) With a white dwarf mass of
$0.24\msun$, the neutron star mass would come down to $1.5\msun$.

We should mention that the existence of a $2.2\msun$ neutron star
would wreak havoc with our scenario of nearly equal masses in the
more massive of the binary pulsars. The most abundant first-born
neutron stars of mass $1.1 -1.2\msun$ from $\sim $ ZAMS $10\msun$
giant progenitors would accrete $\sim 0.5\msun$ in hydrogen red
giant phase and another $0.1 -0.2\msun$ in helium red giant. They
would then end up with pulsar mass $\sim 1.8\msun$, with most
likely low-mass $1.1-1.2\msun$ companion. Such massive pulsars
should be strong in the radio and should predominate in number.
None such have been observed.

%xxxxx

%%%%%%%%%%%%%%%%%%%%%%%%%%%%%%%%%%%%%%%%%%%%%%%%%%%%%%%%%%%%%%%%%
\section{Discussion}
\label{sec7}

Although our discussion based on the bottom-up approach supports
the Akaishi et al. scenario of strangeness nuggets, we have
emphasized the Kaplan and Nelson \cite{KaplanNelson} scalar
attraction from the movement towards restoration of the explicitly
broken chiral symmetry in the strange sector, Eqs.~(\ref{eq7}) and
(\ref{eq8}). Of course, this is more straightforward for us to
handle in our mean field approach, than in the strangeness nugget,
although the ratio of $\langle\bar\psi\psi\rangle$ to vector
density $n$ is model dependent. On the other hand, the vector mean
fields such as $V_{K^-}(\omega) =-\frac 13 V_N$ of Eq.~(\ref{eq1})
are straightforward, the $K^-$ having one nonstrange antiquark,
and the constituent quark model should be adequate at $n=3 n_0$ to
describe the $\frac 13$ ratio of $V_{K^-}$ to $V_N$. The vector
mean field $V_N$ at $n=n_0$ has been much used in Walecka theory
in nuclear physics. A value of $V_N(n_0)=270$ MeV is quite modest
in nuclear physics. {\it It should be noted that this value of $3
V_{K^-}(\omega)$ only reaches this after the factor
$(F_\pi/F_\pi^\star)^2$ is included.} There is a clear indication,
not least of all directly from experiment \cite{Metag2004} that a
sliding vacuum; i.e., Brown-Rho scaling must be operative. The
vector meson masses continue to drop, until they go to zero in the
chiral limit, but as the scale (density in this case) increases,
the coupling $g_V^\star$ also decreases. Near the fixed point
$(g_V^\star/m_V^\star)^2$ goes as a constant. We have handled the
scaling crudely by assuming that $g_V$ does not scale up to $n_0$,
for which there is substantial basis \cite{BR2004}, and then
neglect any further scaling in $g_V^\star/m_V^\star$ above $n_0$.
This gives the factor $(F_\pi/F_\pi^\star)^2 = 1.56$ in the medium
dependence we used. In the above way we achieve a rough
consistency with the Harada and Yamawaki \cite{HY:PR} hidden local
symmetry.
Since all of our important results here refer to $n>n_0$, for
these the sliding in coupling cancels that in meson mass, and we
can carry out the calculations with an effective $f_\pi^\star$
such that $(F_\pi/F_\pi^\star)^2=1.56$ replacing $f_\pi$. This was
also foreseen in the Akaishi et al. work. In short, Brown-Rho
scaling essentially increases the zero-density parameters of mean
fields up to the couplings which have already been used for a long
time in Walecka type mean field approaches.

We suggested that mean field would be much more reliable when
treated starting from the VM fixed point and sketched a highly
simplified calculation that reproduces the bottom-up mean field
calculation for kaon condensation. When further developed, the
top-down approach will prove to be more powerful for the
strangeness nugget problem.

It must be admitted that standard chiral perturbation theory
calculations fluctuating from the zero-vacuum have not been
carried out to the order necessary to get from the input
lowest-order $f_\pi$ taken to be equal to $f_K$ in the necessary
places. It is not clear to us that one can obtain a reliable
result from such an endeavor. It seems that such a calculation
must begin in-medium, from $F_\pi^\star\simeq 0.8 F_\pi$. Thus,
the in-medium $f_K^\star$ would not be expected to be larger than
the $f_\pi$ we have used in $m_K^\star/m_K$.

An interesting calculation in chiral perturbation theory was
carried out by Waas and Weise \cite{WaasWeise} and did go to high
enough order to include the range term (they did not however
employ the medium modified $f_\pi^\star$). They used
$\Sigma_{KN}=230$ MeV, substantially smaller than that found in
LGS. Waas and Weise found that iterating the Weinberg-Tomozawa
term increased the attraction, lowering $m_K^\star/m_K$
appreciably. In this way, their approach was midway between our
mean field and the Akaishi et al. approach, which iterated the
attraction by solving in a potential.

In our discussion of the melting of the soft glue in LGS
\cite{BLRS} we found that nucleons changed over to constituent
quarks at $T\sim 120$ MeV whereas the constituent quarks became
current quarks, having lost their dynamically generated masses, at
$T_c({\rm unquenched})=175$ MeV. Although we do not know the value
of $n_c$, and as we noted earlier, in Nambu Jona-Lasinio
$\langle\bar\psi\psi \rangle$ doesn't really go to zero until well
above the $n_c$ for zero bare quark masses; we suggest that at
$u=n/n_0\sim 3$ constituent quarks may be more appropriate
variables than nucleons, although they will still be held together
by the soft glue in nucleons to some extent.

Thus, although $f_K$ may enter in the Kaplan-Nelson term
Eq.~(\ref{eq8}), the vector mean fields are connected only to the
constituent $\bar u$ quark in the $K^-$, and we believe that all
of these calculations can be handled at the constituent quark
level, but with Brown-Rho scaling of the constituent quark masses.
This suggests working around the VM fixed point as stressed in
\cite{mr-dsb04}. Whereas the lowest-order Weinberg-Tomozawa term
brings $m_K^\star/m_K$ down to 0.65 in nuclear matter (it would
not be brought down so far in neutron stars) with a factor of 1.56
it would be brought down more than half way to zero. The second
order terms are necessarily attractive, and will be increased by a
factor of $(F_\pi/F_\pi^\star)^{4}\simeq 2.4$. There may be
saturation in the complete coupled channels calculation, which
includes Pauli blocking, but the contribution of the
Weinberg-Tomozawa terms will obviously be substantially increased
by the medium dependence. On the other hand, the
$\langle\bar\psi\psi\rangle$ for constituent quarks may be smaller
than the value Thorsson et al. \cite{TPL1994} calculated for
nucleons. We believe that the additional attraction from medium
effects on the vector mean fields would be adequate to compensate
for this.

The strangeness nugget presents a remarkable scenario in which
once the $K^-$ meson is introduced into the three remaining
nucleons in $^4$He, after a proton has been kicked out, the three
nucleons collapse around the $K^-$ into the small nugget system.
What else can they do ? All interactions, scalar from the
Kaplan-Nelson term and vector mean fields from Weinberg-Tomozawa
terms are highly attractive, the latter having only a small
isospin dependence through the $\rho$ mean field, so the system
collapses down to an average density $n\sim 3 n_0$. (It is however
difficult to imagine how this can be achieved in standard chiral
perturbative approaches.) The $K^-$ is in a potential about 600
MeV deep at the center, with binding energy of $\sim 250$ MeV,
even more after strangeness condensation in our neutron stars.
Luckily the collapse is halted in the nugget formation, chiefly by
the finite sizes of the interacting objects. In the case of
neutron stars, there is collapse into black holes and we will not
``see" the black hole, neutron-star binaries until gravitational
waves from their mergers are detected.

One issue we have not addressed in this paper is whether the
collapse to black holes occurs without the kaon condensed state
going into a color superconcuting quark matter or after such a
transition. At the moment one can say practically nothing about
this matter. Should the collapse occur from a quark matter rather
than from a kaon-condensed state, then the question as to how the
kaon-condensed state which we predict must occur before the chiral
transition and which may not be a normal Fermi liquid state
develops the pairing instability that turns the matter to a
(color) superconducting state discussed in the literature. As far
as we know this question has not been studied in the literature.

%%%%%%%%%%%%%%%%%%%%%%%%%%%%%%%%%%%%%%%%%%%%%%%%%%%%%%%%%%%%%%%%%
\section{Conclusions}
\label{sec8}

We have argued that strangeness condensation in neutron stars can
be related to the strangeness nuggets found experimentally by
Suzuki et al. \cite{Suzuki04,Suzuki05}. In both cases, binding of
the $K^-$ meson is responsible for the phenomenon. In the neutron
star matter the applicable mean field method gives $\sim 50$ MeV
(or $\sim 20 \%$) more binding to the $K^-$ than in the nugget,
where $m_{K^-}^\star \sim \frac 12 m_K$. In both cases, the medium
dependence in the vector meson mass $m_\rho^\star$ and
$m_\omega^\star$ provide $\sim 50$ MeV of the binding. Our
arguments are admittedly far from rigorous. If this relation can
be put on a more rigorous basis, it will supply a firm support for
out limit of $1.5\msun$ as maximum neutron star mass. It is
remarkable that in Table~\ref{tab1} out of $\sim 40$ measured
masses of compact objects at most two or three of the masses seem
to violate our upper limit. These and other astrophysical issues
will be discussed in depth in a paper in preparation
\cite{BBL2006}.

\section*{Dedication}
We would like to dedicate this paper to Hans Bethe who fearlessly
coauthored {\it ``A Scenario for a large number of low-mass black
holes in the Galaxy"} \cite{BB1994}, the large number resulting
from the low maximum neutron star mass of $1.5\msun$. Gerry Brown
was explaining the idea of kaon condensation to Hans Bethe on a
Saturday morning walk on a trail near Santa Barbara, California.
Hans immediately understood it, ``You mean that you squeeze
electrons into $K^-$-mesons ?"

%%%%%%%%%%%%%%%%%%%%%%%%%%%%%%%%%%%%%%%%%%%%%%%%%%%%%%%%%%%%%%%%%
\section*{Acknowledgments}
We are grateful to Avraham Gal for helpful comments. GEB was
supported in part by the US Department of Energy under Grant No.
DE-FG02-88ER40388. CHL was supported by grant No.
R01-2005-000-10334-0 from the Basic Research Program of the Korea
Science \& Engineering Foundation.

%%%%%%%%%%%%%%%%%%%%%%%%%%%%%%%%%%%%%%%%%%%%%%%%%%%%%%%%%%%%%%%%%%%%%%%%%%%%%%%%

\appendix
%%%%%%%%%%%%%%%%%%%%%%%%%%%%%%%%%%%%%%%%%%%%%%%%%%%%%%%%%%%%%%%%%%%%%%%%%%%%%%%%
%\section{Appendix A: Comparison with Lattice Gauge Simulation}

\section{Appendix: Dense system with  tribaryon-$\bar{K}$ crystals}
\label{appA}

In this Appendix, we construct a crystalized dense system
comprised of  tribaryon-$K^-$ bound states. The assumptions we
make in the construction of the crystal potential are:
\begin{enumerate}
\item
The effective potential of the single tribaryon-$\bar{K}$
(ppn-$\bar{K}$, Isospin=1) proposed by Akaishi et al.
\cite{Akaishi2005} is used; \be
 V(r) = (V_0 + W_0 ) exp [-(r/a)^2]
\ee with $V_0=-702$ MeV,  $W_0=-13$ MeV and a=0.923 fm. By solving
the Klein-Gordon equation, \be ((E-V(r))^2 -m_K^2-p^2]\psi =0,
\label{KG} \ee Akaishi et al. obtained the binging energy and rms
radius of the tribaryon-$\bar{K}$ as 194 MeV and 0.74 fm,
respectively. In this estimate, the enhanced core energy of the
tribaryon from the core shrinkage, $\delta E \simeq 110$ MeV, is
assumed. Akaishi at el. show the average density of the
tribaryon-$\bar{K}$ bound system to be $ n \simeq 3.1 n_0$.

\item We put these tribaryon-$K^-$ systems in a cubic lattice
by assuming that the potential generated by the tribaryon is the
same as that given by Akaishi et al.

\item
In a dense medium, a tribaryon-$\bar{K}$ bound system is affected
by the surrounding tribaryon cores. In order to take this effect
into account, we make an effective potential of a single site by
making the average over all solid angle. Finally we have an
effective potential with spherical symmetry.

\end{enumerate}

With the above assumptions, we obtain the binding energies and the
rms radii of the tribaryon-$K^-$ system. In order to take into
account the density effects, we change the distances between these
tribaryon-$K^-$ systems.

\begin{figure}[t]
\centerline{\epsfig{file=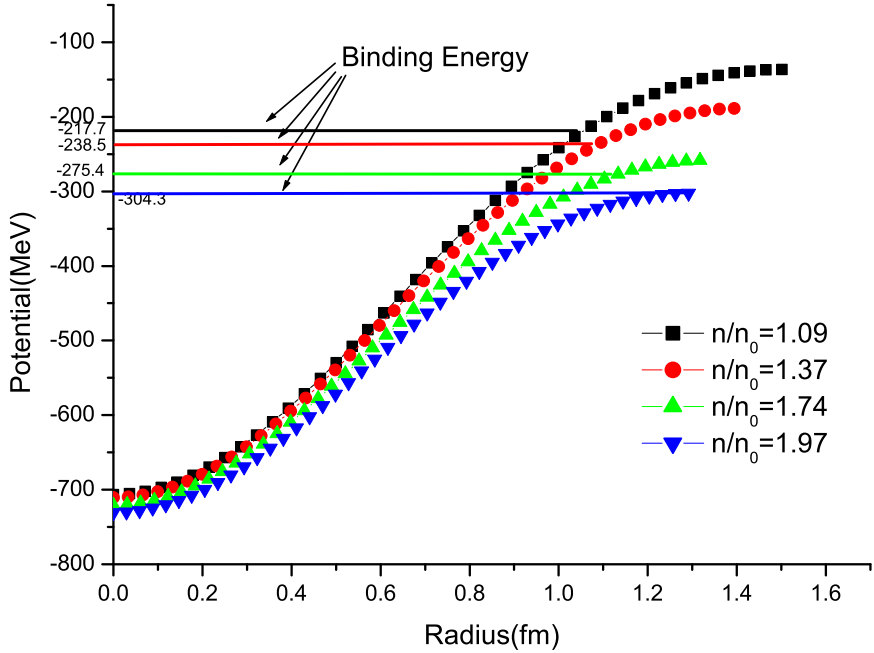,height=4in}}
\caption{Effective potentials for various baryon densities.}
\label{potential}
\end{figure}

In Figure~\ref{potential}, we see that the minimum and the
threshold of the effective potential of a tribaryon-$K^-$ system
are lowered as the baryon density increases due to the overlap of
the potential given by the neighboring systems. As one can see
from the figure, when the binding energy is equal to the maximum
value of the potential, the bound state of the tribaryon-$\bar{K}$
system dissolves.

\begin{table}[t]
\begin{center}
\begin{tabular}{l|lllll}
\hline\hline
  $ n/n_0 $  &  1.09  &  1.37  & 1.74 &   1.97&  2.02    \\
\hline
B.E.(MeV)     &    -217.7  &   -238.5   &  -275.4  &  -304.3    &    -310.7   \\
$ V_{max} $     &    -136.6  &   -188.3  &  -258.1 &  -302.2   &   -311.9  \\
rms Radius(fm) &    0.816 &   0.842  &  0.862 &   0.868  &  unbound    \\

\hline\hline

\label{BEnergy}
\end{tabular}
\caption{Binding Energy and rms radius of a single
tribaryon-$\bar{K}$ system in medium for various densities. Beyond
$2n_0$, the bound tribaryon-$K^-$ system ``dissolves" indicating
that the the crystalized system is unstable.}
\end{center}
\end{table}

As in Table~\ref{BEnergy}, the bound state of the
tribaryon-$\bar{K}$ disappears at $ n/n_0 \simeq 2.0$. This
indicates that for densities greater than $2 n_0$, the system
becomes continuous, losing the crystal structure. We interpret
this as kaon condensation taking place from $2 n_0$. Since we
started with the fixed proton fraction (1/3) in this approach, the
density at which the crystal dissolves cannot be directly compared
with the critical density for kaon condensation in the mean field
approach. However, our results provide a way to link
tribaryon-$K^-$ bound states (nuggets) to kaon condensation.

%%%%%%%%%%%%%%%%%%%%%%%%%%%%%%%%%%%%%%%%%%%%%%%%%%%%%%%%%%%%%%%%%%%%%%%%%%%%%%%%


\begin{thebibliography}{99}
\bibitem{TPL1994} V. Thorsson, M. Prakash, and J.M. Lattimer,
Nucl. Phys. {\bf A 572} (1994) 693.

\bibitem{BB1994}
G.E. Brown and H.A. Bethe, Astrophys. J. {\bf 423} (1994) 659.

\bibitem{BB1995}
H.A. Bethe and G.E. Brown, Astrophys. J. {\bf 445} (1995) L129.

\bibitem{Brown2001}
G.E. Brown, A. Heger, N. Langer, C.-H. Lee, S. Wellstein, and H.A.
Bethe, New Astronomy {\bf 6} (2001) 457.

\bibitem{BWW}
G.E. Brown, J.F. Weingartner, and R.A.M.J. Wijers, Astrophys. J.
{\bf 463} (1996) 297.

\bibitem{BBL2006}
H.A. Bethe, G.E. Brown, and C.-H. Lee, in preparation (2006).

\bibitem{BR2004}
G.E. Brown and M. Rho, Phys. Reps. {\bf 396} (2004) 1.

\bi{BR2002} G.E. Brown and M. Rho, Phys. Reps. {\bf 363} (2002)
85; B. Friman and M. Rho, Nucl. Phys. {\bf A606} (1996) 303.

\bi{shankar} See, e.g., R. Shankar, Rev. Mod. Phys. {\bf 66}
(1994) 129.

\bibitem{HY:PR} M. Harada and K. Yamawaki, Phys. Rep. {\bf 381} (2003) 1.

\bibitem{HRS} M. Harada, M. Rho and C. Sasaki, Phys.Rev. {\bf D70}
(2004) 074002.

\bibitem{mr-dsb04} M. Rho, ``Pentaquarks, skyrmions and the vector
manifestation of chiral symmetry," hep-ph/0502049.



\bibitem{kaonrot} G.E. Brown, V. Thorsson, K. Kubodera and M. Rho,
Phys. Lett. {\bf B291} (1992) 355.

\bibitem{Dong1996}
J.J. Dong, J.-F. Lagae, and K.F. Liu, Phys. Rev. {\bf D 54} (1996)
5496.

\bibitem{BRwalecka} G.E. Brown and M. Rho, Nucl.Phys. {\bf A596} (1996)
503.

\bibitem{Suzuki2004}
K. Suzuki et al., Phys. Rev. Lett. {\bf 92} (2004) 072302.

\bibitem{BR91} G.~E.~Brown and M.~Rho,
%``Scaling effective Lagrangians in a dense medium,''
Phys.\ Rev.\ Lett.\  {\bf 66} (1991) 2720.

\bibitem{Metag2004}   D. Trnka et al [CBELSA/TAPS Collaboration],
``First observation of in-medium modifications of the omega
meson," nucl-ex/0504010, to appear in Phys. Rev. Lett.

\bibitem{SBMR} C. Song, G.E. Brown, D.-P. Min and M. Rho, Phys. Rev.
{\bf C56} (1997) 2244.

\bibitem{Song}
C. Song,  Phys. Rept. {\bf 347} (2001) 289.

\bibitem{Suzuki04}
A. Dot\'e, H. Horiuchi, Y. Akaishi, and T. Yamazaki,
Phys. Lett. {\bf B 590} (2004) 51.


\bibitem{Akaishi2005}
Y. Akaishi, A. Dot\'e, and T. Yamazaki, ``Strange tribaryons as
$\overline{K}$-mediated dense nuclear systems," nucl-th/0501040.

\bibitem{Mares2}
J. Mares, E. Friedman and A. Gal, Phys. Lett. {\bf B606} (2005)
295.

\bibitem{LBMR} C.-H. Lee, G.E. Brown, D.-P. Min and M. Rho, Nucl.Phys. {\bf A585} (1995)
 401.

\bibitem{Lee:PR}
C.-H. Lee, Phys. Rept. {\bf 275} (1996) 255.

\bibitem{Bernard87} V. Bernard, U.G. Meissner, and I. Zahed,
                   Phys. Rev. D {\bf 36} (1987) 819.
\bibitem{BGLR} G.E. Brown, L. Grandchamp, C.-H. Lee and M. Rho,
Phys. Rept. {\bf 391} (2004) 353.

\bibitem{Lutz}
M. Lutz, S. Klimt, and W. Weise, Nucl. Phys. {\bf A 542} (1992)
521.

\bibitem{Suzuki05}
T. Suzuki et al., ``A search for deeply bound kaonic nuclear
states," nucl-ex/0501013.

\bibitem{BLS}
G.E. Brown, M. Rho, and C. Song, Nucl. Phys. {\bf A 690} (2001)
184c.

\bibitem{HY:PRL} M. Harada and K. Yamawaki, Phys. Rev. Lett. {\bf 86} (2001) 757.

\bibitem{KochBrown93} V. Koch and G.E. Brown, Nucl. Phys. {\bf A 560} (1993) 345.

\bibitem{tsushima} K. Tsushima, K. Saito, A.W. Thomas and S.V.
Wright, Phys. Lett. {\bf B429} (1998) 239.

\bibitem{KV2003}
E.E. Kolomeitsev and D.N. Voskresensky, Phys. Rev. {\bf C 68}
(2003) 015803.

\bibitem{BFG}
C.J. Batty, E. Friedman, and A. Gal, Phys. Rept. {\bf 287} (1997)
386.

\bibitem{BLRapp1998}
G.E. Brown, C.-H. Lee, and R. Rapp, Nucl. Phys. {\bf A 639} (1998)
455c.

\bibitem{Saha}
P.K. Saha et al., Phys. Rev. {\bf C70} (2004) 044613.

\bibitem{Noumi}
H. Noumi et al., Phys. Rev. Lett. 89 (2002) 072301; erratum Phys.
Rev. Lett. {\bf 90} (2003) 049902.

\bibitem{Harada}
T. Harada and Y. Hirabayashi, NPA, to appear.

\bibitem{BFG94}
C.J. Batty, E. Friedman, A. Gal, Phys. Lett. {\bf B335} (1994)
273.

\bibitem{Mares}
J. Mares, E. Friedman, A. Gal, B.K. Jenning, Nucl. Phys. {\bf
A594} (1995) 311.

\bibitem{vdH1993}
E.P.J. Van den Heuvel and J. van Paradijs, 1993,
Scientific American, November (1993) 38.

\bibitem{BB1998}
H.A. Bethe and G.E. Brown, Astrophys. J. {\bf 506} (1998) 780.

\bibitem{BKB}
K. Belczynski, V. Kalogera, and T. Bulik, Astrophys. J. {\bf 572}
(2002) 407.


\bibitem{Brown1995}
G.E. Brown, Astrophys. J. {\bf 440} (1995) 270.

\bibitem{BraunLanger}
H. Braun and N. Langer, Astron. Astrophys. {\bf 297} (1995) 483.

\bibitem{Burrows}
A. Burrows and S.E. Woosley,  Astrophys. J. {\bf 308} (1986) 680.

\bibitem{paradijs2}
J. van Paradijs, E.J. Zuiderwijk, R.J. Takens, and G.
Hammerschlag-Hensberge, Astron. Astrophy. Supplementary 30 (1977)
195.

\bibitem{barziv}
O. Barziv, L. Kaper, M.H. van Kerkwijk, J.H. Telting, J. Van
Paradijs, Astron. Astrophy. {\bf 377} (2001) 925.

\bibitem{MKKH} A. van der Meer, L. Kaper, M.H. van Kerkwijk,
and E.P.J. van den Heuvel, Proceedings of the International
Workshop ``Massive Stars in Interacting Binaries", La maison du
lac Sacacomie, Canada, Aug. 16-20, 2004; astro-ph/0502313.

\bibitem{KaplanNelson}
D.B. Kaplan and  A.E. Nelson, Phys. Lett. {\bf B 175} (1986) 57.

\bibitem{WaasWeise}
T. Waas and W. Weise, Nucl. Phys. {\bf A 625} (1997) 287.

\bibitem{BLRS} G.E. Brown, C.-H. Lee, M. Rho, and E.V. Shuryak,
Nucl. Phys. {\bf A 740} (2004) 171.


%\bibitem{dote} A. Dot\'e, H.Horiuchi and Y. Akaishi, nucl-th/0309062.

\end{thebibliography}
\end{document}